\documentclass[prl,
twocolumn,
 amsmath,amssymb,
 aps,
]{revtex4-2}

\usepackage{graphicx}
\usepackage{dcolumn}
\usepackage{bm}
\usepackage{braket}

\DeclareMathOperator\arctanh{arctanh}
\usepackage{hyperref}
\hypersetup{colorlinks,urlcolor=blue,linkcolor=red,          
citecolor=blue,        
filecolor=blue}

\begin{document}


\title{Fate of entanglement in magnetism under Lindbladian or non-Markovian dynamics and conditions for their transition to  Landau-Lifshitz-Gilbert classical dynamics}

\author{Federico Garcia-Gaitan}
\affiliation{Department of Physics and Astronomy, University of Delaware, Newark, DE 19716, USA}
\author{Branislav K. Nikoli\'{c}}
\email{bnikolic@udel.edu}
\affiliation{Department of Physics and Astronomy, University of Delaware, Newark, DE 19716, USA}


\begin{abstract}
The entanglement of many localized spins (LS) within 
solid magnetic materials is a topic of great basic and applied interest, particularly after becoming amenable to experimental 
scrutiny where, e.g., very recent neutron scattering experiments have  witnessed macroscopic entanglement in the ground state (GS) of antiferromagnets persisting even at elevated temperatures. On the other hand, spintronics and magnonics studies assume that LS of  antiferromagnets are in unentangled  N\'{e}el GS, as well as that they evolve, when pushed out of equilibrium by current or external fields, according to the Landau-Lifshitz-Gilbert (LLG) equation viewing LS as classical vectors of {\em fixed} length. The prerequisite for applicability of the LLG equation is {\em zero entanglement} in the underlying many-body quantum state of LS. In this study, we initialize quantum Heisenberg ferro- or antiferromagnetic chains hosing  \mbox{$S=1/2$}, \mbox{$S=1$} or \mbox{$S=5/2$} LS into unentangled pure state and then evolve them by quantum master equations (QMEs) of Lindblad or non-Markovian type, derived by coupling LS weakly to bosonic bath (due to  phonons in real materials) or by using additional ``reaction coordinate'' in the latter case. The time evolution is initiated by applying an external magnetic field, and entanglement of the ensuing {\em mixed} quantum states is monitored by computing its negativity. We find that non-Markovian dynamics {\em never} brings entanglement to zero, in the presence of which the vector of spin expectation value {\em changes its length} to render the LLG equation {\em inapplicable}. Conversely, Lindbladian (i.e., Markovian) dynamics ensures that entanglement goes to zero, thereby enabling quantum-to-classical transition in all cases---$S=1/2$, $S=1$ and $S=5/2$  ferromagnet  or $S=5/2$ antiferromagnet---{\em except} for $S=1/2$ and $S=1$ antiferromagnet. Finally, we  investigate the stability of entangled antiferromagnetic GS upon suddenly coupling it to  bosonic baths.
\end{abstract}

\maketitle
{\em Introduction.}---The fate of entanglement of many interacting quantum spins, localized at the sites of  crystalline lattices of magnetic materials~\cite{Christensen2007} or in optical lattices of their quantum  simulators~\cite{Brydges2019}, under finite temperature or nonequilibrium conditions is a topic of great contemporary interest. For example, recent experiments~\cite{Scheie2021, Mathew2020, Laurell2021}  have succeeded to witness~\cite{Friis2018,Chiara2018, Laflorencie2016} multipartite entanglement~\cite{Scheie2021,Song2011} of macroscopically large number of spins hosted by  antiferromagnetic insulators (AFIs) in equilibrium up to surprisingly high temperature \mbox{$T \lesssim 200$ K}~\cite{Scheie2021}. Transient entanglement in nonequilibrium AFIs could also be witnessed via very recently proposed schemes~\cite{Hales2023,Baykusheva2023}. Due to finite temperature and/or nonequilibrium, such systems inevitably generate {\em mixed entangled} states, also in the focus of our study [Eq.~\eqref{eq:mixed_unentangled}], that are far less understood~\cite{Peres1996,Wu2020,Elben2020,Sang2021,Aolita2015} than the pure~\cite{Brydges2019} entangled ones. In computational quantum physics, quantum spin systems are a standard playground for developing algorithms, such as tensor networks (TN)~\cite{Banuls2023}, that can  efficiently encode ground states (GSs) containing low-entanglement---however, entanglement growth in nonequilibrium~\cite{Bardarson2012} poses significant challenge for these algorithms~\cite{Trivedi2022} and the role of dissipative environment in limiting the so-called ``entanglement barrier'' is intensely explored~\cite{Lerose2023}. It is insightful to invoke a pedagogical example of an entangled GS, such as that of AFI chain hosting localized spins (LS) $S=1/2$, which has been realized experimentally~\cite{Sahling2015} and is described by the Heisenberg Hamiltonian~\cite{Essler2005} \mbox{$\hat{H}_H = J \sum_{i = 1}^{N-1} \hat{\mathbf{S}}_{i}\cdot \hat{\mathbf{S}}_{i+1}$}. The GS is entangled~\cite{Song2011} as it cannot be expressed as the direct product of multiple single-spin states in any basis, as obvious from its form for $N=4$ sites: 
$
|\mathrm{GS}\rangle_\mathrm{AFI}  =   \frac{1}{\sqrt{12}} \big( 2\ket{\uparrow \downarrow \uparrow \downarrow} + 2\ket{\downarrow \uparrow \downarrow \uparrow} - \ket{\uparrow\uparrow \downarrow  \downarrow}
 - \ket{\uparrow \downarrow \downarrow \uparrow  }  - \ket{\downarrow \downarrow \uparrow \uparrow } -\ket{\downarrow \uparrow \uparrow \downarrow} \big).
$
Its energy, \mbox{$_{\mathrm{AFI}}\langle \mathrm{GS}| \hat{H} | \mathrm{GS} \rangle_{\mathrm{AFI}} = -2J$}, is lower  than the energy, \mbox{$\langle \mathrm{N\acute{e}el} | \hat{H} | \mathrm{N\acute{e}el}\rangle=-J$} of unentangeled  N\'{e}el state $\ket{\mathrm{N\acute{e}el}} = \ket{\uparrow \downarrow \uparrow \downarrow}$, which is the precise meaning behind ``quantum spin fluctuations''~\cite{Singh1990} sintagma.  Here \mbox{$\hat{S}_{i}^\alpha =\hat{I}_1 \otimes \ldots \otimes S \hat{\sigma}^\alpha \otimes \ldots \otimes \hat{I}_\mathrm{N_\mathrm{AFI}}$} acts nontrivially, as the Pauli matrix $\hat{\sigma}^\alpha$, in the Hilbert space of spin at   site $i$; $\hat{I}_i$ is the unit operator; and  $J > 0$ is AF exchange interaction. The expectation value of spin, $\langle \hat{\mathbf{S}}_i \rangle = \langle \mathrm{GS}|\hat{\mathbf{S}}_i |\mathrm{GS} \rangle \equiv 0$, vanishes  as a direct consequence~\cite{Wieser2015,Mondal2021,Petrovic2021b} of non-zero  entanglement entropy of AFI GS.  

In the case of ferromagnetic insulators (FIs),  quantum spin fluctuations~\cite{Singh1990} are absent~\cite{Pratt2004} and both classical $\uparrow \uparrow \ldots \uparrow \uparrow$ and its unentangled quantum counterpart 
$\ket{ \uparrow \uparrow \ldots \uparrow \uparrow }$ are GS of the respective classical and quantum Hamiltonians. However, excited states of FI chain---such as one-magnon Fock state~\cite{Bajpai2021, Yuan2022} 
\mbox{$|1_q \rangle = \frac{1}{\sqrt{N}}\sum_{n = 0}^{N-1}e^{iqx_n}|  \underbrace{ \!\uparrow\ldots\!\uparrow}_\text{$n$}\downarrow   \underbrace{\!\uparrow\ldots\!\uparrow}_\text{$N-n-1$}\rangle$} 
where  $q$ is the wavevector and \mbox{$x_n=n a$} is the $x$-coordinate  along the chain (with the lattice constant $a$)---is macroscopically entangled~\cite{Morimae2005,LachanceQuirion2020},  as is the case of multi-magnon states~\cite{Pratt2006}. The robustness of entanglement of such states has been studied for a long time in quantum computing (using analogous multi-qubit states known as $W$ states)~\cite{Carvalho2004,Aolita2015}, as well as more recently in ``quantum magnonics''~\cite{Yuan2022} using quantum master equations (QMEs) formulated in second-quantization~\cite{Yuan2022a,Yuan2022b}.
 
On the other hand, it is commonly assumed in antiferromagnetic spintronics~\cite{Baltz2018,Jungwirth2016,Zelezny2018,Jungfleisch2018,Gray2019} that GS of AFI is unentangled N\'{e}el state;  or that excited states (like magnons~\cite{Ritzmann2020,Suresh2021}) of either AFI or FI, as triggered experimentally by injected current~\cite{Cheng2014,Gray2019,Wang2019a} or electromagnetic radiation~\cite{Vaidya2020,Li2020,Qiu2021}, are {\em classical and governed}~\cite{Cheng2014,Yuan2019} by the celebrated Landau-Lifshitz-Gilbert (LLG) equation~\cite{Landau1935,Gilbert2004,Saslow2009}. It is also widely-believed that large spin value $S$~\cite{Wieser2015} and/or room temperature ensure applicability of the LLG equation. This plausible notion is motivated by the  eigenvalue of $\hat{\mathbf{S}}_i^2$ operator being $S^2(1+1/S)$, instead of $S^2$, which suggests that quantum effects become progressively less important for $S>1$. 
However, even for single quantum spin the required value of $S$ to match quantum and classical LLG dynamics can be unrealistically large~\cite{Gauyacq2014,GarciaPalacios2006} in the presence of magnetic anisotropy (or any quadratic or higher order terms in spin Hamiltonian)~\cite{Wieser2013}. Also, quantum corrections persist for all $S < \infty$~\cite{Gauyacq2014,GarciaPalacios2006}, vanishing as $(2S)^{-1}$ in the classical limit~\cite{Parkinson1985}. Importantly, most of the standard magnetic materials host LS with rather small $S \le 5/2$~\cite{Kaxiras2019}. 

The search for a rigorous proof that quantum dynamics of a {\em single} spin can transition to classical LLG dynamics, due to interaction with dissipative environment like the bosonic bath and conditions imposed on it, has a long history dating back to archetypical spin-boson model~\cite{Leggett1987} and recent generalizations~\cite{Anders2022} completing the proof while also unraveling nature of quantum corrections to classical LLG dynamics. However, such  proofs~\cite{Anders2022} do not explain how quantum dynamics of {\em many} spins can transition to classical dynamics to be describable by a system of coupled LLG equations~\cite{Evans2014}, often applied without scrutiny to both ferro- and antiferromagnets in spintronics~\cite{Cheng2014} and magnonics~\cite{Kim2010}. The {\em key prerequisite} for such a transition is the {\em absence of entanglement}~\cite{Wieser2015,Mondal2021}, i.e., the underlying quantum state of many LS must remain unentangled pure
\mbox{$
\ket{\sigma_1(t)}\otimes\ket{\sigma_2(t)} \otimes \dots \otimes \ket{\sigma_N(t)}
$}, or unentangled mixed~\cite{Peres1996,Wu2020,Elben2020,Sang2021,Aolita2015} 
\begin{equation}
\label{eq:mixed_unentangled}
\hat{\rho}(t) = \sum_n p_n \hat{\rho}_n^{(1)}(t) \otimes \hat{\rho}_n^{(2)}(t) \dots \hat{\rho}_n^{(N)}(t),
\end{equation}
at all times $t$ in order for time evolution of quantum-mechanical expectation values $\langle \hat{\mathbf{S}}_i \rangle$ to be able to transition to the solutions~\cite{Evans2014}  $\mathbf{S}_i(t)$ of  coupled LLG equations
\begin{equation}
    \langle \hat{\mathbf{S}}_i \rangle(t) \mapsto \mathbf{S}_i(t).
\end{equation}
Otherwise, in the entangled quantum state the length of vectors $\langle \hat{\mathbf{S}}_i \rangle (t)$ is changing in time~\cite{Petrovic2021a} which obviously {\em cannot be mimicked} by $\mathbf{S}_i(t)$ of {\em fixed length}~\cite{Evans2014} in the LLG equation. In Eq.~\eqref{eq:mixed_unentangled}, $\hat{\rho}_n^{(i)}$ is the density matrix spin at site $i$.  We consider usage of the LLG equation in the context of atomistic spin dynamics (ASD)~\cite{Evans2014}, where each atom of the lattice hosts one classical vector $\mathbf{S}_i$.

\begin{figure*}[t!]
    \centering
    \includegraphics[width=\textwidth]
    {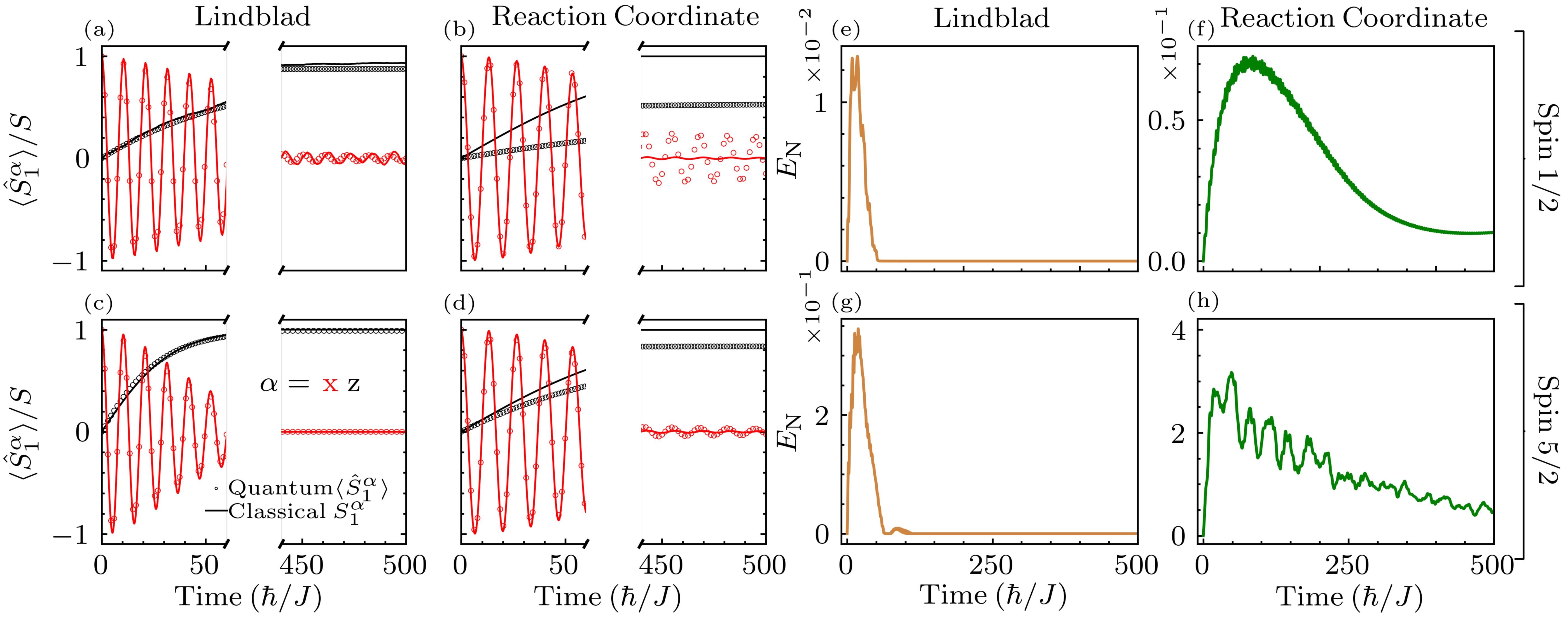}
    \caption{Time-dependence of QME-computed $\langle \hat{S}_1^\alpha \rangle(t)$ vs. classical-LLG-computed  $S_1^\alpha(t)$ of LS (a),(b) $S=1/2$ or (c),(d) $S=5/2$ on site $i=1$ of FI chain of $N=4$ sites. (e)--(h) Time dependence of entanglement negativity~\cite{Peres1996} $E_N(t)$ between two halves of FI chain in the case of quantum evolution (circles) in panels (a)--(d). The LS open quantum system evolution is conducted using either Lindblad (i.e., Markovian) or RC (i.e., non-Markovian) QME with bosonic bath temperature  $T=|J|$.}
    \label{fig:FM_sim}
\end{figure*}

\begin{figure*}
    \centering
    \includegraphics[width=\textwidth]{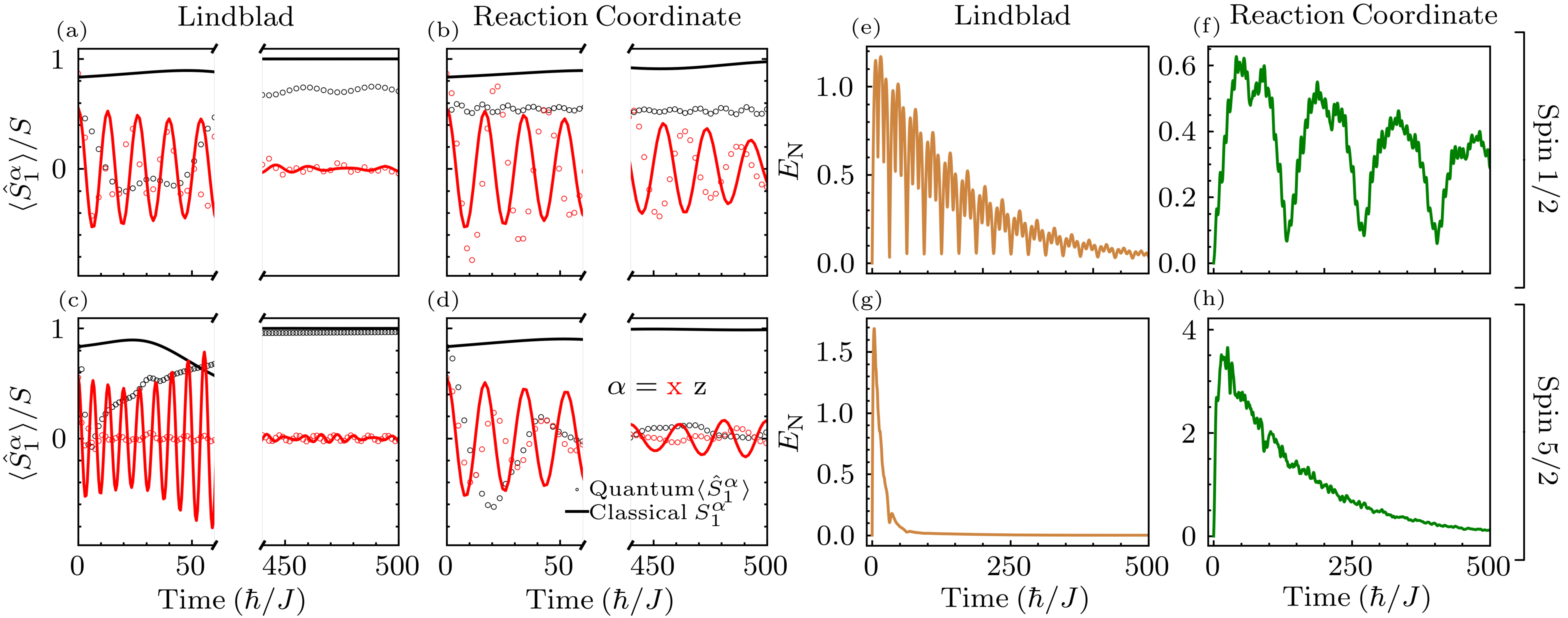}
    \caption{Panels (a)--(h) are counterparts of Fig.~\ref{fig:FM_sim}(a)--(h), but using AFI chain composed of  $N=4$ sites. Additional cases of AFI or FI chains hosting $S=1$ LS, or by including their additional interactions (such as easy-axis anisotropy, long-ranged  dipole-dipole interaction and Dzyaloshinskii-Moriya interaction) are provided in the SM~[74].}
    \label{fig:AFM_sim}
\end{figure*}

\begin{figure}
    \centering
    \includegraphics[scale=0.15]
    {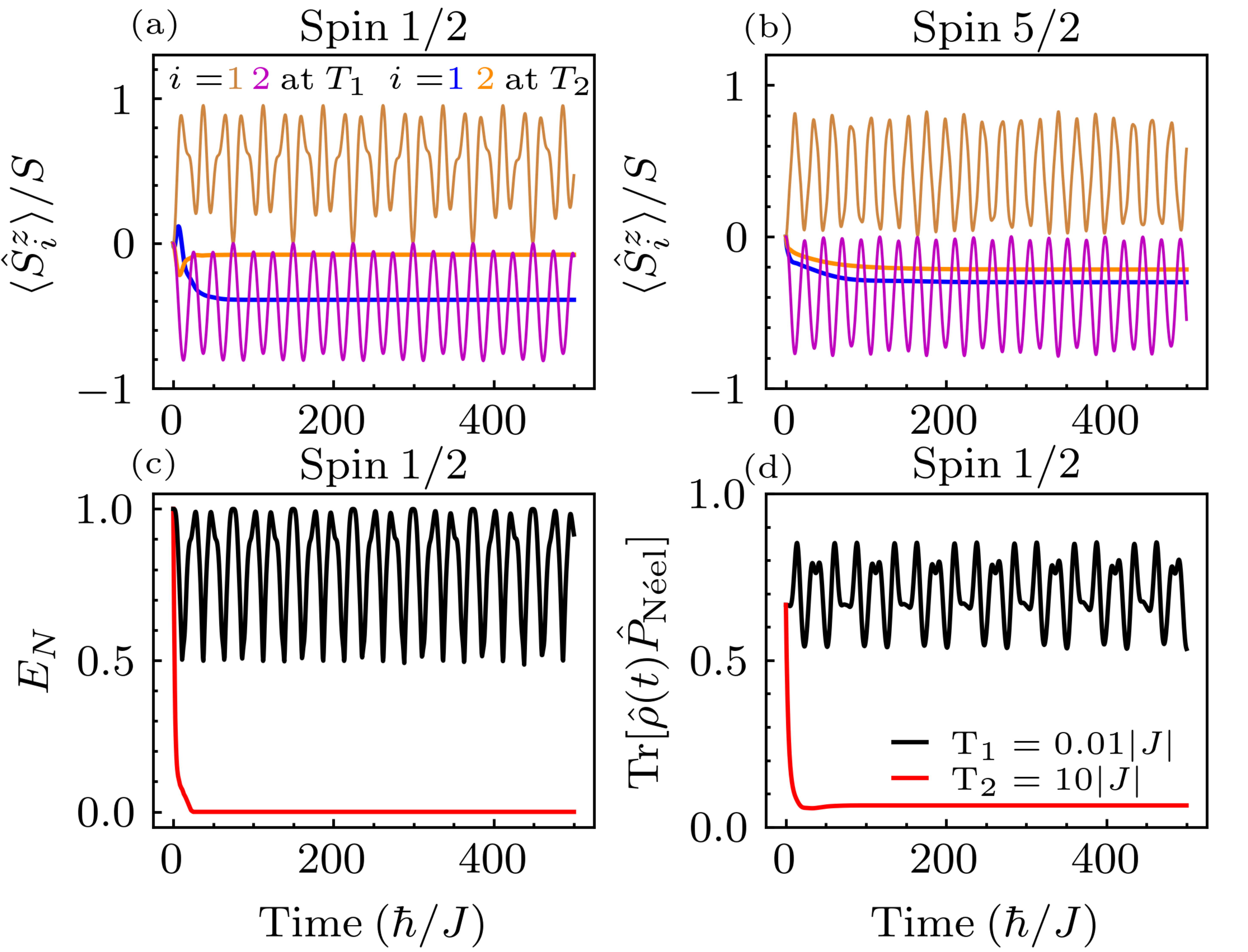}
    \caption{Time dependence of: (a),(b) spin expectation values at sites $i=1,2$; (c) entanglement negativity $E_N(t)$~\cite{Peres1996} between two halves of AFI chain; and (d) overlap between the chain density matrix $\hat{\rho}(t)$ and pure states in the N\'{e}el subspace. The AFI chain has $N=4$ sites, as well as an impurity introducing the $z$-axis anisotropy at site $i=1$ [Eq.~\eqref{eq:himp}]. The Lindblad Eq.~\eqref{eq:ULE}  evolves $\hat{\rho}(t)$ upon coupling AFI chain to the bosonic bath at $t=0$, starting from pure entangled GS but exhibiting N\'{e}el ``checkerboard'' order $\langle \hat{S}_{i}^z \rangle = - \langle \hat{S}_{i+1}^z \rangle \neq 0$~\cite{Petrovic2021b}.} 
    \label{fig:neel}
\end{figure}

In this Letter, we view AFI and FI as open quantum systems~\cite{Breuer2007,Breuer2016,Vega2017} by coupling them either: ({\em i}) weakly to a bosonic bath, assumed to arise due to bosonic quasiparticles in solids such as phonons, whose tracing out allows one to derive~\cite{Rudner2020} the universal Lindblad QME [Eq.~\eqref{eq:ULE}]; or ({\em ii}) strongly to a single bosonic mode which, in turn, interacts weakly with the bosonic bath, so that tracing over both allows us to derive non-Markovian QME  within the so-called ``reaction coordinate'' (RC) method~\cite{Schaller2014}.  We monitor the presence of entanglement in the density matrix of all LS $\hat{\rho}(t)$ via entanglement negativity $E_N(t)$~\cite{Peres1996,Wu2020, Elben2020, Sang2021,Aolita2015}, and we concurrently compare quantum $\langle \hat{\mathbf{S}}_i \rangle(t)$ vs. classical  $\mathbf{S}_i(t)$ trajectories in Figs.~\ref{fig:FM_sim}--\ref{fig:neel}. 


{\em Models and methods.}---We consider FI ($J<0$) or AFI ($J>0$) chain modelled by the Heisenberg Hamiltonian $\hat{H}_H$, which can include interaction with a homogeneous external magnetic field switching on for $t\geq 0$, \mbox{$\hat{H} = \hat{H}_H - \sum_i g\mu_B\mathbf{B}_{\mathrm{ext}}(t\geq 0) \cdot \hat{\mathbf{S}}_i$}, where $g$ is the gyromagnetic ratio and $\mu_B$ is the Bohr magneton. We set $\hbar=1$ and $k_B=1$. These  models of realistic magnetic materials~\cite{Scheie2021,Sahling2015} are made open quantum systems by coupling them bosonic baths, so that the total Hamiltonian becomes 
$
    \hat{H}_{\text{tot}}= \hat{H}+\hat{H}_{\text{bath}}+\hat{V}.
$
Here $\hat{H}_{\text{bath}}$ models a set of independent baths, one per each spin~\cite{Rudner2020,Weber2022}, as harmonic oscillators~\cite{Leggett1987}, $\hat{H}_{\text{bath}}= \sum_{ik} w_{ik} \hat{a}_{ik}^\dagger \hat{a}_{ik}$, using an operator $\hat{a}_{ik}(\hat{a}_{ik}^\dagger)$ which annihilates (creates) a boson in mode $k$. The boson interacts with spin operator at site $i$~\cite{Breuer2007} via $\hat{V} = \sum_k g_k \sum_i  \hat{\mathbf{S}}_{i}  (\hat{a}_{ik} + \hat{a}_{ik}^\dagger)$,  where $g_k$ are the coupling constants. By assuming small $g_k$, QME of the Lindblad type~\cite{Manzano2020,Lindblad1976} can be derived by tracing out the bosonic bath and by expanding the resulting equation to second order. Rather than relying on traditional approaches for the derivation of the Lindblad QME---such as using Born, Markov and secular approximations~\cite{Manzano2020, Schaller2014,Norambuena2020}---we follow the procedure of Ref.~\cite{Rudner2020} for universal Lindblad QME which evades  difficulties of the secular approximation~\cite{Schaller2016}. For example, for systems with (nearly) degenerate eigenstates, as is the case of FI and AFI models considered, secular approximation leads to improperly derived~\cite{Norambuena2020} Lindblad QME for LS because of assuming that energy splitting is much bigger than fluctuations due to the bath. The same problem was addressed in a number of recent studies~\cite{Lidar2020, McCauley2020}, besides the resolution offered in Ref.~\cite{Rudner2020}.

The universal Lindblad QME~\cite{Rudner2020} considers a single Lindblad operator $\hat{L}_i$ for each spin, so that only $N$ such operators are needed to obtain
 \begin{equation}
    \label{eq:ULE}
    d{\hat{\rho}}/dt = -i [\hat{H},\hat{\rho}] + \sum_i^N \hat{L}_i \hat{\rho} \hat{L}^\dagger_i-\frac{1}{2}\{\hat{L}_i^\dagger \hat{L}_i, \hat{\rho} \},
\end{equation}
where we also ignore typically negligible Lamb-shift corrections~\cite{Schaller2014} to the Hamiltonian. The Lindblad QME is time-local due to the assumption that bath-induced changes to the system dynamics are slow relative to the typical correlation time of the bath. We compute $\hat{L}_i$ operators as a power series (where we use cutoff  $N_L \le 20$)
\begin{equation}
    \label{eq:expand}
    \hat{L}_i = \sum_{n=0}^{N_L} c_n (\text{ad}_{\hat{H}})^n [\hat{\mathbf{S}}_i], \: \: \: \: c_n=\frac{(-i)^n}{n!}\int_{-\infty}^\infty dt g(t) t^n,
\end{equation}
thereby evading the need for exact diagonalization~\cite{Rudner2020} of FI or AFI Hamiltonians. Here $\text{ad}_{\hat{H}} [X] = [\hat{H},X]$ and the jump correlator function is defined via the Fourier transform of the spectral function of the bath, \mbox{$J(\omega)=2\pi \sum \delta(\omega-\omega_k)$}, as
    \mbox{$g(t)=\frac{1}{\sqrt{2\pi}}\int_{-\infty}^{\infty} d\omega\sqrt{J(\omega)} e^{-i \omega t}$}.
For numerical calculations, we considered an Ohmic~\cite{Anders2022} spectral function with a rigid ultraviolet cutoff
\begin{equation} \label{eq:Spectral}
    J(\omega) = \frac{\Gamma \omega}{\omega_m}  n_{\mathrm{BE}}(\omega)\Theta(\omega_m-\omega),
\end{equation}
where $\Gamma$ is the reorganization energy representing the magnitude of fluctuations and dissipation; $\omega_m$ characterizes how quickly the bath relaxes towards equilibrium; $n_\mathrm{BE}(\omega)$ the Bose-Einstein distribution; and $\Theta$ is the Heaviside step function. 

The Lindblad QME [Eq.~\eqref{eq:ULE}] is only valid for a weak system-bath coupling, as it assumes a second order truncation in $g_k$. Since this is not always the case, several approaches~\cite{Breuer2007,Vega2017} exist to treat strong system-bath coupling, such as polaron, star-to-chain and thermofield transformations~\cite{Landi2022}, and the RC  method~\cite{Schaller2016}. The RC method we employ is based on the Bogoliubov transformation, and it allows one to construct a new bosonic mode $\hat{b}$ called the RC. This mode is coupled strongly to the system, but weakly to a residual bosonic bath, while conserving the bosonic commutation relations. The new Hamiltonian of the system then becomes
\begin{equation}
    \label{eq:RCHamiltonian}
    \hat{H}_{\text{tot}} = \hat{H} + \lambda \sum_i \hat{\mathbf{S}}_i  (\hat{b} + \hat{b}^\dagger) + \Omega \hat{b}^\dagger \hat{b} + \hat{H}_{\text{RC-B}} + \hat{H}_{\text{bath}},
\end{equation}
where $\lambda$ is the strength of the coupling between the RC and the system; $\Omega$ is the frequency  of the RC; $\hat{H}_{\text{RC-B}}=\sum_{k>1} \tilde{g}_k (\hat{b} + \hat{b}^\dagger) (\hat{c}_k + \hat{c }_k^\dagger)$ is the RC-bath coupling Hamiltonian; and $\hat{H}_{\text{bath}}$ is the bosonic bath Hamiltonian considered to be identical to the case used in derivation of Eq.~\eqref{eq:ULE}, but with one less bosonic mode and with properly transformed coupling coefficients. Thus, the parameters $\lambda$ and $\Omega$ are expressed~\cite{Nazir2018} in terms of the parameters in Eq.~\eqref{eq:Spectral},
$\lambda ^ 2 = \frac{1}{6\pi} \sqrt{\frac{5}{3}} \Gamma \omega_m$ and $\Omega = \sqrt{\frac{5}{3}} \omega_m$, while the spectral function of the residual bath 
\begin{widetext}
\begin{equation} \label{eq:SpectralRC}
    J^\prime(\omega)= \frac{2 \sqrt{5/3} \pi \omega \omega_m^2n_{\mathrm{BE}}(\omega)}{3 \left[ \pi^2 \omega^2 +4 \omega \arctanh(\omega/\omega_m) (\omega \arctanh (\omega/\omega_m) -2\omega_m) +4\omega_m^2\right]},
\end{equation}
\end{widetext}
is independent of the original coupling strength $\Gamma$. This allows us to derive QME which has the same form as Eq.~\eqref{eq:ULE},  but it uses \mbox{$\hat{H} \mapsto \hat{H} + \lambda \sum_i \hat{\mathbf{S}}_i  (\hat{b} + \hat{b}^\dagger) + \Omega \hat{b}^\dagger \hat{b}$}. Since \mbox{$\lambda \propto \sqrt{\Gamma}$}, the coupling of the system to the RC can be arbitrarily strong without affecting coupling to the residual bath. Despite being time-local, this Lindblad QME including  RC captures non-Markovian effects~\cite{Segal2021}. They, otherwise, require integro-differential QMEs with time-retarded kernel~\cite{Breuer2007,Breuer2016,Vega2017}. In order to reduce computational complexity for many LS, an effective Hamiltonian was built by considering~\cite{Segal2021} only the lowest energy states of the RC, i.e., the matrix representation of $\hat{b}$  is truncated to finite size $15 \times 15$.

{\em Results and discussion.}---We solve Eq.~\eqref{eq:ULE} for Lindbladian dynamics, as well as for non-Markovian dynamics when the RC is included in the Hamiltonian, for FI and AFI chains composed of $N=4$ sites with periodic boundary conditions hosting spins $S=1/2$ or  $S=5/2$, as well as $S=1$ in the Supplemental Material (SM)~\footnote{See Supplemental Material at \href{https://wiki.physics.udel.edu/qttg/Publications}{https://wiki.physics.udel.edu/qttg/Publications}, which includes  Refs.\cite{GarciaPalacios1998, Zou2020, Davis2023, sbierski2023, iemini2023, Prosen2014, Queisser2019, Seif2022, lee2020, nielsen_chuang_2010, Cresser2021, Nathan2022, Szilva2023, Camley2023, Erickson1992, Sharma2001a, Schuckert2018}, for: additional Fig.~S1 as the counterpart of Figs.~\ref{fig:FM_sim} and ~\ref{fig:AFM_sim}, but considering $S=1$ LS within either FI or AFI chains; numerical procedures for solving the LLG equation for $\mathbf{S}_i(t)$, as well as for comparing $\mathbf{S}_i(t)$ with $\langle \hat{\mathbf{S}}_i \rangle(t)$ which also allows one to extract {\em microscopcially} the Gilbert damping in the LLG equation; the effect of additional interactions acting on LS, such as easy-axis anisotropy, dipole-dipole interaction and Dzyaloshinskii–Moriya interaction; usage of a single vs. many independent bosonic baths; exploration of wide range of temperatures of bosonic baths; proximity of steady-state solution of the Lindblad QME to Gibbs density matrix describing thermal equilibrium; and scaling of entanglement with increasing number of spins.}. The two QMEs are solved using the fourth order Runge-Kutta method, where $|J|=1$ sets the unit of energy. For Lindbladian dynamics we use $\Gamma =0.01 |J|$, while for non-Markovian dynamics we use stronger coupling $\Gamma=0.1|J|$, and the cutoff frequency is chosen as $\omega_m=3|J|$. Note that choosing too large $\omega_m$ brings entanglement of LS to zero on a very short timescale. The initial condition for FI is unentangled pure state \mbox{$\hat{\rho}(0)= |\Sigma \rangle \langle \Sigma|$}, where \mbox{$|\Sigma\rangle = \ket{ \rightarrow \rightarrow \rightarrow \rightarrow }$}  with all spins pointing along the $x$-axis. The magnetic field applied for $t \ge 0$, $g\mu_B B_z = 0.8 |J|$, is along the $z$-axis. The initial condition for AFI is unentangled pure state $\hat{\rho}(0)= |\Omega \rangle \langle \Omega|$, where $|\Omega\rangle= |\sigma_1 \sigma_2 \sigma_1 \sigma_2 \rangle$ with $\langle \sigma_{1(2)}|\hat{\mathbf{S}}_{1(2)}|\sigma_{1(2)}\rangle$ pointing along $\theta_1=1/8$ or $\theta_2=\pi-1/8$ and $\phi_{1(2)}=0$ in spherical coordinates.

In the course of time evolution, $\hat{\rho}(t)$ can become entangled which is quantified by computing entanglement negativity~\cite{Peres1996,Wu2020, Elben2020, Sang2021,Aolita2015} between the left half (LH) and the right half (RH) of the chain 
		 $E_{N}[\hat{\rho}(t)] = \ln || \hat{\rho}^{T_{\mathrm{RH}}} ||_1  = \ln \sum_n |\lambda_n|$ ,
where  $||\hat{A}||_1 =\mathrm{Tr} \sqrt{\hat{A}^\dagger \hat{A}}$ is the trace norm of the operator $\hat{A}$; $\lambda_n$ are the eigenvalues of $\hat{\rho}^{T_{\mathrm{RH}}}$; and the matrix elements of the partial transpose with respect to RH  of the chain are given by $\big(\hat{\rho}^{T_{\mathrm{RH}}}\big)_{i\alpha;j\beta} = \big(\hat{\rho} \big)_{j\alpha;i\beta}$. While the standard von Neunmann entanglement entropy $\mathcal{S}_{\mathrm{LH}}$ of half of the chain~\cite{Bardarson2012,Chiara2018} can be non-zero even for unentangled mixed state in Eq.~\eqref{eq:mixed_unentangled}, non-zero $E_N$ necessarily implies entanglement and genuine quantum correlations between the two parts~\cite{Peres1996,Elben2020, Wu2020, Sang2021, Aolita2015}.  

Initially, both FI and AFI exhibit dynamical build-up of entanglement signified by $E_N>0$ in Figs.~\ref{fig:FM_sim} and ~\ref{fig:AFM_sim}, respectively. However, Lindbladian dynamics quickly brings $E_N \rightarrow 0$ in the FI hosting $S=1/2$ [Fig.~\ref{fig:FM_sim}(e)], $S=1$ (Fig.~S1(g) in the SM~\footnotemark[1]), and $S=5/2$ [Fig.~\ref{fig:FM_sim}(g)] spins; as well as in the AFI hosting $S=5/2$ [Fig.~\ref{fig:AFM_sim}(e)] spins. Establishing $E_N \rightarrow 0$ also makes it possible for LLG classical  trajectories $\mathbf{S}_i(t)$ to track $\langle \hat{\mathbf{S}}_i \rangle(t)$ in Figs.~\ref{fig:FM_sim}, ~\ref{fig:AFM_sim} and S1 in the SM~\footnotemark[1]. Details of how the LLG equation is solved to obtain $\mathbf{S}_i(t)$, while tuning the Gilbert damping parameter in order to enable comparison of  $\mathbf{S}_i(t)$ and $\langle \hat{\mathbf{S}}_i \rangle(t)$, are given in the SM~\footnotemark[1]. In the AFI case with $S=1/2$ [Fig.~\ref{fig:AFM_sim}(e)] or $S=1$ (Fig.~S1(c) in the SM~\footnotemark[1]) entanglement {\em never} vanishes, $E_N(t)>0$, even in the long-time limit, thereby maintaining $\langle \hat{\mathbf{S}}_i \rangle(t) \neq \mathbf{S}_i(t)$. Thus, we conclude that usage~\cite{Baltz2018,Jungwirth2016,Zelezny2018,Jungfleisch2018,Cheng2014} of the LLG equation in spintronics with AFI layers hosting spins $S=1/2$ or $S=1$ {\em cannot be justified microscopically}. In the case of non-Markovian dynamics, $E_N(t)$ remains non-zero (Figs.~\ref{fig:FM_sim}, ~\ref{fig:AFM_sim} and S1 in the SM\footnotemark[1]) in FI and AFI at all times and for $S=1/2$, $S=1$ and $S=5/2$, so that quantum-to-classical transition $\langle \hat{\mathbf{S}}_i \rangle(t) \mapsto \mathbf{S}_i(t)$ is {\em never} achieved. This then provides an example of how pronounced memory effects can lead to revival of genuine quantum properties such as quantum coherence, correlations, and entanglement~\cite{Breuer2016}.

Finally, we examine the fate of entangled GS of AFI upon {\em suddenly coupling} it to a bosonic bath and evolving it by the Lindblad Eq.~\eqref{eq:ULE}. Let us recall that a common trick employed in TN calculations on spin systems to select the unentangled N\'{e}el state as the GS is to introduce an external staggered magnetic field which alternates in sign on atomic length scales~\cite{Stoudenmire2012}. However, its microscopic justification is missing. Attempts to introduce more realistic decoherence mechanisms, such as repeated local measurements~\cite{Katsnelson2001,Donker2016,Donker2018} that would disrupt superposition in the GS and replace the need for contrived staggered field, are also difficult to justify in the context of spintronic and magnonic devices. A handful of recent studies have examined time evolution of entangled GS of AFIs~\cite{Weber2022,Schaller2022} upon suddenly coupling their spins to a dissipative environment, but with conflicting conclusions about the fate of entanglement. 
Since the ``checkerboard'' pattern of expectation values of $\langle \hat{\mathbf{S}_i}\rangle$ in the N\'{e}el order is often reported  experimentally~\cite{Gray2019}, we induce it as the initial condition at $t=0$ by using GS of slightly modified Heisenberg Hamiltonian
\begin{equation}\label{eq:himp}
\mbox{$\hat{H}_\mathrm{imp} = \hat{H}_H- 0.2 |J|\hat{S}_1^z$}, 
\end{equation}
with an additional impurity at site $i=1$. The impurity  breaks rotational invariance of $\hat{H}_H$  to  generate N\'{e}el order, $\langle \hat{S}_{i}^z \rangle = - \langle \hat{S}_{i+1}^z \rangle \neq 0$, {\em but not} the N\'{e}el GS $|\!\! \uparrow\downarrow \uparrow\downarrow \rangle$ because entanglement entropy of true GS remains nonzero~\cite{Petrovic2021b} leading to $\langle \hat{S}_{i}^z \rangle/S<1$. The Lindbladian time evolution [Fig.~\ref{fig:neel}] maintains entanglement $E_N(t)>0$ at low temperature $T_1=0.01|J|$ and, therefore, {\em nonclassical}  dynamics of $\langle \hat{\mathbf{S}}_i \rangle(t)$, while at high temperatures $E_N \rightarrow 0$ is reached on short time scales. The overlap $\mathrm{Tr}\, [\hat{\rho}(t) \hat{P}_\mathrm{N\acute{e}el}]$ with states in the  N\'{e}el subspace, whose projector is $\hat{P}_\mathrm{N\acute{e}el}=\ket{\uparrow\downarrow \uparrow\downarrow}\bra{\uparrow\downarrow \uparrow\downarrow} + \ket{ \downarrow \uparrow \downarrow \uparrow} \bra{  \downarrow \uparrow \downarrow \uparrow}$, never reaches 1 in the low temperature regime [black curve in Fig.~\ref{fig:neel}(d)]. In the high temperature limit, the overlap becomes negligible [red curve in Fig.~\ref{fig:neel}(d)] as the system goes~\cite{Schaller2022} into {\em static ferrimagnetic} ordering [blue and orange flat lines in Figs.~\ref{fig:neel}(a) and ~\ref{fig:neel}(b)]. 

In conclusion, we solve nearly a century old~\cite{Landau1935} problem---``unreasonable effectiveness'' of the classical LLG equation in describing dynamics of {\em many} (for solution of 
the same problem for a single spin, see Ref.~\cite{Anders2022}) localized spins within a magnetic material---by showing that it is justified microscopically only if Lindblad open quantum system dynamics is generated by environment in the case of any ferromagnet, as well as for antiferromagnets with sufficiently large value of their spin $S>1$. {\em Thus, our findings exclude  antiferromagnets with $S=1/2$ or $S=1$ spins from possibility to model them via  classical micromagnetics or ASD}~\cite{Kim2010,Evans2014}. Our analysis via rigorously constructed Markovian and non-Markovian QMEs for many 
LS interacting with bosonic bath could also be applied to other related problems, such as fate of entanglement in quantum spin liquids~\cite{Yang2021}. 

\begin{acknowledgements}
This research was primarily supported by the US National Science Foundation through the University of Delaware Materials Research Science and Engineering Center, DMR-2011824.
\end{acknowledgements}

\nocite{*}
\bibliography{biblio}

\end{document}